# Thermal regimes of Li-ion conductivity in β-eucryptite

Yachao Chen[1], Sukriti Manna[2], Cristian V. Ciobanu,[2*] and Ivar E. Reimanis[1*]

[1]Department of Metallurgical and Materials Engineering, Colorado School of Mines, Golden, CO 80401, USA

[2]Department of Mechanical Engineering, Colorado School of Mines, Golden, CO 80401, USA

**Abstract:** While it is well established that ionic conduction in lithium aluminosilicates proceeds via hopping of Li ions, the nature of the various hoping-based mechanisms in different temperature regimes has not been fully elucidated. The difficulties associated with investigating the conduction have to do with the presence of grains and grain boundaries of different orientations in these usually polycrystalline materials. Herein, we use electrochemical impedance spectroscopy (EIS) to investigate the ion conduction mechanisms in β-eucryptite, which is a prototypical lithium aluminosilicate. In the absence of significant structural transitions in grain boundaries, we have found that there are three conduction regimes for the one-dimensional ionic motion along the *c* axis channels in the grains, and determined the activation energies for each of these temperature regimes. Activation energies computed from molecular statics calculations of the potential energy landscape encountered by Li ions suggest that at temperatures below 440 °C conduction proceeds via cooperative or correlated motion, in agreement with established literature. Between 440 °C and 500°C, the activation barriers extracted from EIS measurements are large and consistent with those from atomistic calculations for uncorrelated Li ion hopping. Above 500 °C the activation barriers decrease significantly, which indicates that after the transition to the Li-disordered phase of β-eucryptite, the Li ion motion largely regains the correlated character.

Corresponding authors. Email: cciobanu@mines.edu (CVC) and reimanis@mines.edu (IER)

## 1. Introduction

As a prototype of lithium aluminum silicates (LAS), β-eucryptite (LiAlSiO$_4$) has attracted both fundamental and technological interest for decades due to its phase transformations,[1-5] unusual thermo-mechanical properties,[6-10] and the one-dimensional nature of its ionic conduction.[11-15] The superior Li ion conduction makes β-eucryptite a promising candidate for electrolyte applications, including thermal batteries and high-temperature solid electrodes.[16-18] Furthermore, applications such as electrothermal devices as well as thermal shock resistant structures can benefit from the overall negative coefficient of thermal expansion (CTE) of β-eucryptite.[8,9,16,17,19-24] The crystal structure of β-eucryptite (space group P6$_4$22 or P6$_2$22)[1,25-29] is a Li-stuffed derivative of β-quartz in which half of the Si-centered tetrahedra, [SiO$_4$]$^{4-}$, are replaced with Al-centered tetrahedra, [AlO$_4$]$^{5-}$.[30,31] In ordered β-eucryptite, [SiO$_4$]$^{4-}$ and [AlO$_4$]$^{5-}$ tetrahedra are arranged in a spiral along the 6$_4$ (or 6$_2$) screw axis, forming open channels along which the Li ions reside at well-defined locations.[1,20] The effect of temperature on this structure consists in disordering of the Li ions at ~440 °C,[27,30,32-37] while at very high temperatures disordering of the Si- and Al- centered tetrahedra could also occur.[20,30]

Li motion in the channels is of special interest[8,38] for applications of eucryptite as an ionic conductor. It has been reported that the Li ionic conductivity parallel to the *c*-axis is three orders of magnitude greater than that perpendicular to the *c*-axis in the temperature range from 200 °C to 600 °C,[12,13] which makes β-eucryptite an excellent material to study one-dimensional diffusion.[33,34,39] While such studies have been reported in the literature for different preparations and different temperature regimes[12-15,21,34,39-41] and the mechanism of conduction is accepted to be Li ion hopping along the channels, there is a rather large range of activation energies reported for the hopping. For example, the activation energy for hopping below 500 °C has been reported to be 0.89 eV,[15] 0.8 eV,[21] 0.79 eV,[41] 0.74 eV,[12,13] and 0.62 eV,[39] with the variations attributed to differences in the measurement or computation techniques, crystallinity, and sample preparations. The variation from 0.62 eV[39] to 0.89 eV[15] is significant, as the same temperature would lead to vastly different conductivity values for these energy values because of the activated nature of Li ion hopping. The presence of channels in the structure of eucryptite may lead one to expect large ionic conductivity and low activation energy. However, the activation energy of β-eucryptite is not particularly small, as illustrated by any of the above figures.[12-15,21,34,39-41] Two factors may conspire to decrease the ionic conductivity in polycrystalline eucryptite: first, under normal processing conditions, the grains in the microstructure are randomly oriented and therefore their fast conducting *c*-axis channels are not all parallel to each other; second, grain boundaries likely present an impediment to Li-ion diffusion.

Despite a significant body of work on Li-ion conduction in β-eucryptite,[12,14,16,18,41,42] particularly on single-crystal and glasses, there remain two key outstanding questions for the polycrystalline material. First, why is there a variation of activation energy over a large temperature range? Such variation may correspond to different mechanisms operating in different thermal regimes, or it may correspond to the (continuous) thermal expansion or contraction of the structure along different directions. Second, what is the role of the conduction through grain boundaries and other slow ion conduction pathways (such as hopping perpendicular to the *c*-axis)? Using electrochemical impedance spectroscopy (EIS) measurements performed at temperatures up to 900 °C and atomistic-based modeling, we address these questions here and show that thermal expansion/contraction does not play a signficant role in determining the mechanism of ionic conductivity. We determined the activation energies at different temperatures based on a brick layer model analysis of the EIS data, and identified three temperature regimes characterized by constant (and different) energy barriers. At temperatures below 440 °C, conduction proceeds via cooperative or correlated motion, in agreement with established literature. Between 440 °C and 500°C, the activation barriers extracted from EIS measurements are consistent with those computed from atomistic calculations for uncorrelated Li ion hopping. Above 500 °C the activation barriers obtained from EIS decrease significantly, which indicates that after the transition to the Li-disordered phase of β-eucryptite, the Li ion motion largely regains the correlated character. The correlated diffusion of Li atoms has often referred to as superionic conductivity. The main conclusion drawn from the present results is that as temperature is increased from room temperature through the order-disorder transformation temperature, there is a loss (at 440 -500 °C) and then a re-entry (past 500 °C) of the superionic character of the conduction.

## 2. Experimental Procedure

Pure β-eucryptite powders were synthesized via a chemical precursor route,[23,43,44] in which tetraethylorthosilicate (TEOS-Sigma Aldrich, St. Louis, MO USA), aluminum nitrate ($Al(NO_3)_3 \cdot 9H_2O$-Sigma Aldrich), and lithium nitrate ($LiNO_3$-Sigma Aldrich) were used as precursors. Proper amounts of ethanol (95%), distilled water, and $HNO_3$ were added to help form a homogeneous solution. This solution was then treated with excess ammonium hydroxide (~10-15%) to form gel, which was then dried at 80 °C to obtain amorphous powders. These powders were first heated up to 400 °C and held for 1 hour in order to remove water and gaseous products formed by decomposition of the nitrates, then calcined in air at 1100 °C for 15 hours to form single phase β-eucryptite as confirmed by x-ray diffraction. The crystalline β-eucryptite powders were then ball-milled using $ZrO_2$ balls and ethanol to achieve a final particle size of about 1 μm.[23] The powders were subsequently poured into a graphite die, 25 mm in diameter, and sintered in a vacuum hot press at 1200 °C for 2 h under an applied load of 30 MPa, followed by slow cooling (1°C/min).[23] All pellets were found to have about 97%–98% of the theoretical density (2.34 g/cm$^3$) and remained single phase β-eucryptite.

The samples used for the ionic conductivity measurements were cut with a diamond saw to obtain the dimensions of about 1 × 1 × 0.1 cm$^3$ and were polished on the surfaces to be contacted. Silver mesh and gold wire was attached to the pellet surfaces using platinum as a current collector. Electrochemical impedance spectroscopy (EIS) of β-eucryptite was performed by a two-probe method using Solartron with a signal amplitude of 10 mV under open circuit voltage (OCV) conditions in the temperature range of 300 °C to 900 °C with the frequency from 1 Hz to 6MHz. A thermocouple was placed next to the sample in the tube. Data analysis was performed with software Zview (Scribner Associates, Southern Pines, NC).[45] A schematic microstructure based on a brick layer model (BLM)[46] are shown in Figure 1, which illustrates fast conduction portions (green, conduction parallel to the $c$ axis) and slow portions (grain boundaries and unfavorably oriented grains) for a given pathway.

## 3. Results and Discussion

Impedance spectroscopy coupled with the BLM model for analysis[47] is a technique meant strictly for a "composite" medium with different ionic conductivities for the two isotropic components that make up the sample. For example, the use of this technique for polycrystalline zirconia[48,49] yields both the conductivity of the grains (high) and of the grain boundaries (low). In the case of zirconia, the grains are isotropic or very nearly so, and the grain boundaries are assumed isotropic. Applying EIS for highly anisotropic materials such as β-eucryptite poses several problems: (i) the conduction pathways pass through grains that do not all conduct equally well because of their $c$ axes are not aligned (Fig. 1); (ii) grain boundaries are slow conduction avenues, and as such they would act similar to the slow-conducting (i.e., unfavorably oriented) grains. Therefore, the conductivity and the activation barriers obtained from EIS measurements performed on highly anisotropic materials have an approximate or "effective" character: the parallel conduction (and the associated barrier) refers to pathways that are parallel or nearly parallel to *c* axis, while the slow conduction necessarily includes both grain boundaries and unfavorably oriented grains. Below we describe our results and analysis in terms of sample resistances determined from EIS, ionic conductivities, and activation energies for Li conduction processes.

Figure 2(a) displays several typical impedance spectra of pure β-eucryptite at selected temperatures between 300 °C to 900 °C in Nyquist form, with the negative imaginary part ($-Z_{imag}$) plotted as a function of the real part ($Z_{real}$). As mentioned, the analysis of polycrystalline materials is more complicated than that for single crystals because of the presence of grain boundaries. For modeling isotropic materials, the equivalent circuits normally adopt the elements corresponding to bulk, grain boundary, and electrode.[46,50,51] However, in highly anisotropic materials, such as β-eucryptite, the effect of different grain orientations is not completely captured by the simple equivalent circuits because the grain boundaries and slow conducting grains cannot be readily resolved. In general, the capacitance values per unit length are in the range of $10^{-11}$ to $10^{-8}$ F/cm for the grain boundaries and of the order of $10^{-12}$ F/cm for bulk.[52] The anisotropy is very large for β-eucryptite, *i.e.* the conductivity along the *c*-axis is about a thousand times higher than that perpendicular; the capacitance associated with both directions are similar,[52] so the EIS response of the unfavorably oriented grains overlaps largely with that of the grain boundaries. The simplified equivalent circuit, therefore, adopts two parallel RC circuits connected in series (Figure 1(b)). In Figure 1(b), $R_\parallel$ and $CPE_\parallel$ represent the conduction along the *c*-axis in the favorably oriented grains; $R_{slow}$ and $CPE_{slow}$ represent the slow paths and correspond to arcs in the Nyquist plots for temperatures between 300 °C and 440 °C in Figure 2. The sample-electrode impedance is represented by $CPE_{electrode}$.

Inspection of the data presented in Figure 2 reveal that there are two types of impedance plots, one consisting only of a linear portion (Figure 2(b)), and the other comprising an arc and a linear portion at low frequencies (Figure 2(c)). Figure 2(b) represents the case when $R_{slow}$ decreases and becomes comparable twith $R_{\|}$ (e.g., at temperatures higher than 440 °C); in such cases, it is possible to extract only the total resistance ($R_t$) from the intercept of the linear portion with the horizontal axis.[47,53] Figure 2(c) corresponds to the case where the characteristic frequency associated with the Li motion along the c axis is within the experimentally attainable frequency domains: this corresponds to temperatures from 300 °C to 440 °C, where the Nyquist plots consist of an arc followed by linear spike at lower frequencies. The low-frequency tails in the spectra are due to the blocking effect of the platinum electrode,[54-56] and they are well described using a constant-phase element. The high-frequency end of the arc does not pass through the origin when extrapolated. The non-zero intercept at high frequencies corresponds to the grain bulk resistance for fast conduction (i.e. parallel to the c axis), $R_{\|}$. The equivalent circuit is shown as an inset in Figure 2(c). A second resistance, $R_{slow}$, that effectively corresponds to grain boundaries and unfavorably oriented grains, is extracted from the length of the horizontal chord of the arc, as illustrated in Figure 2(c). The resistance of the contact electrodes is negligible, so $R_t = R_{\|} + R_{slow}$.

The resistance determined from the Nyquist plots can be used to determine the overall (total) conductivity of the samples using $\sigma = L/RA$, where A is the electrode cross section area and L is the length of the pellet; for temperatures below 440 °C, we can use length and area estimates to also determine the effective conductivity of the slow paths (grain boundary and unfavorably oriented grains) and that along the c axis in grains. These conductivity results are plotted in Figure 3. The values in Figure 3 for the one dimensional conductivity $\sigma_{\|}$ are consistent with those reported by Alpen et al,[12,13] and those for the total conductivity σ$_t$ are somewhat lower than those from Shin-Ici et al.[14]

Next, we determine and discuss activation energies, which determine ionic conductivity via the Arrhenius equation

$$\sigma T = B \exp\left(-\frac{E_a}{kT}\right), \qquad (1)$$

in which is the conductivity, $B$ is a pre-exponential constant, $E_a$ is the activation energy, $k$ is Boltzmann's constant, and $T$ is the temperature. This relation holds for the one dimensional bulk conduction along the $c$ axis. In an effective or average way, we will adopt this Arhennius relation for the slow paths as well, since both in the grain boundary and perpendicular to the $c$ axis the Li ion diffusion is more likely to occur by hoping rather than by partial attack of Li ions on the Si-O bonds.[57] Expressing the conductivity in terms of resistance ($\sigma = L/RA$), subsequently taking the natural logarithm of Equation (1), leads to a linear relation between $\ln R/T$ and $1/kT$ with the slope equal activation energy ($-E_a$):

$$\ln\frac{R}{T} = \ln\frac{L}{BA} + \frac{E_a}{kT} \qquad (2)$$

The temperature dependence of $\ln(R/T)$ is shown in Figure 4. In processing the data shown in Figure 4, we have assumed that the motion along slow pathways (grain boundaries, and motion across the c axis channels) remains governed by the same mechanisms and therefore its effective activation energy does not change. We have therefore determined the corresponding activation energy $E_{slow}$ from EIS data at temperatures below 440 °C, and then used this value to compute the resistance $R_{slow}$ for all temperatures above 440 °C. From these extrapolated $R_{slow}$ values and the total resistance $R_t$ (from EIS measurements above 440 °C), we have extracted the one-dimensional (along *c* axis) resistance as $R_\parallel = R_t - R_{slow}$ at all the temperatures for which it could not be obtained directly from Nyquist plots. Figure 4 shows three distinct regions for the one-dimensional ionic conductivity in the bulk; also, by virtue of our initial assumption, there is only one region for the slow conduction along the grain boundaries and across the *c*-axis.

The activation energies obtained from using Equation 2 for fitting various line segments in Figure 4 are listed in Table I for different temperature regimes. A sudden change in activation energy for the one-dimensional conduction ($Bulk_\parallel$ in Table I) in the temperature range from 440 °C to 500 °C is apparent. However, the activation energy at high-temperatures (500 °C to 900 °C) and low-temperature range (300 °C to 440 °C) are relatively similar.

**Table I.** Activation energies of the total, bulk, and grain boundary in β-eucryptite, determined from resistance measurements using Arrhenius's law. Above 440 °C, the activation energy is assumed to be 1.197 eV, hence no standard deviation is reported at these temperatures.

|  | Activation Energy (eV) | |
| --- | --- | --- |
| Temperature range (°C) | *Bulk*$_\parallel$ | *Slow path* |
| 900-500 | 0.573±0.015 | 1.197 |
| 500-440 | 1.750±0.072 | 1.197 |
| 440-300 | 0.455±0.03 | 1.197±0.014 |

The physical origin of the different temperature regimes for the activation barriers is investigated using atomic scale calculations in which Li ion pathways on the potential energy landscape in a *c*-axis channel have been traced using the ReaxFF potential.[58] The ReaxFF potential was previously used to study the amorphisation of β-eucryptite under pressure[4,5] and its response to irradiation.[5,59] The diffusion barriers have been determined via molecular statics calculations, in which Li ions are moved along one channel in a 2×2×1 supercell while the total energy is evaluated at each position along that channel. The results are shown in Figure 5, in which two cases were considered: one where all Li ions along one channel moved in unison (Figure 5(a)), and the other in which only one Li ion was moved along the channel from one local minimum of the energy to the next (Figure 5(b)). These are referred to as correlated and uncorrelated diffusion, respectively.[21] Figure 5 shows that the barrier for correlated Li motion (~0.4 eV) per Li atom is significantly smaller than the energy required for uncorrelated hopping (~1.2 eV). These results are obtained from single crystal calculations with a semi-empirical interatomic potential.[58] As such, they are qualitatively consistent with the activation barriers for the diffusion along *c*-axis (bulk$_\parallel$) obtained from the EIS experiments (Table 1): in both of ReaxFF calculations and the experimental values, there is a factor of ~3 between the activation energy associated with the uncorrelated and correlated motion. Regardless of the correlated or uncorrelated hopping, the activation energy values from the ReaxFF calculations are somewhat smaller than those obtained in experiments: however, this does not prevent the assignment of correlated and uncorrelated motion to various temperature regimes, since the difference between barriers is rather large. Using ReaxFF, we have carried out the same calculations eucryptite crystals with various lattice constants corresponding to different temperatures up to 900 °C: for these cases, the energy barriers determined via ReaxFF have not registered significant variations with temperature, which is consistent with the low thermal expansion coefficient of β-eucryptite.[6-10] We conclude that the thermal expansion of the lattice does not affect the diffusion mechanism in any of the three regimes identified in Figure 4.

We now focus on providing a qualitative reasoning for the relative magnitudes of the activation energies listed in Table 1. Those are intrinsically related to the order-disorder transition in β-eucryptite, in which Li ions change their positions in the channels in the temperature range from 460 °C to 550 °C.[1,3,30,34,60] At room temperature, Li atoms occupy half of the available sites in the channels. There are two types of channels (one primary and three secondary per unit cell), with six available tetrahedral sites in each channel. Li ions in the primary channel occupy sites with $z/c = 1/6, 1/2, 5/6$ along the origin channel ($x = y = 0$), and are coplanar with Al atoms. Li ions occupy the other three secondary channels on sites with $z/c = 0, 1/3, 2/3$, and are coplanar with Si atoms. This type of distribution occurs because the Al tetrahedra sheets are thicker than the Si tetrahedra sheets, since Al ions are larger than Si ions. Stuffing Li ions into within the Si layers reduces the dimensional mismatch of the Si sheets with Al sheets by enlarging the Si-tetrahedra sheets.[20] As a result, Li is distributed such that 25% of the channel Li ions are coplanar with the Al sheet and the other 75% are coplanar with the Si sheet for a fully ordered β-eucryptite at room temperature.[20] At temperatures above the transition, Li ions interchange with the initial vacancy sites, resulting in a random distribution over all framework channel sites.[27] This breaks the 25 to 75 ratio of Al tetrahedra sheets to Si tetrahedra sheets and causes the shrinkage of the size of the Si tetrahedral sheets. Thus, the activation energy for conduction in the disordered structure is slightly higher than that at low temperature for the ordered structure. Since there is no significant change in framework structure before and after the onset of the order-disordered transformation, the activation energies of order and disordered structure remain similar (Table I). However, we determined that between the ordered and disordered structure (440 °C to 500 °C), β-eucryptite has a very high activation energy (Table I). This likely occurs because during the transformation itself the Li ions are not interspersed regularly through the structure as dictated by Coulumb interactions, but are transiting from the Si sheets to the Al sheets while at times hoping into neighboring interstices (vacancies). An example of what hoping into adjacent vacancy sites leads to is shown in Figure 2b: in this figure, the energy barrier is very mainly because the final position (right hand side of the graph) of the moving ion is too close to another Li ion. While we do not simulate the full order-disorder transformation, it is very likely that some Li ions hop too close to one another during disordering transformation,[30] leading to the high barrier reported in the intermediate temperature regime in Table I. A similar phenomenon has been detected in solid acid proton conductor $CsH_2PO_4$, where the disordered structure at high temperature brings the conductivity to a much higher level while the activation energy remains similar to the low temperature ordered structure.[53]

## 4. Conclusion

In conclusion, we have reported on the conductivity of polycrystalline β-eucryptite in various temperature regimes, focusing on resistance, ionic conductivity, energy barriers and the associated hopping mechanisms. Under the assumption that the grain boundaries do not change their structure significantly, we have extracted the one-dimensional conductivity along *c*-axis channels in polycrystalline samples, which has only been done previously for single crystals.[12,13] Although the conduction along grain boundaries and that perpendicular to *c*-axis cannot be individually resolved in our EIS experiments, the premise of a common effective or average activation energy yields a value of 1.197 eV for the slow pathways. While not atypical, this value is rather high, and shows that grain boundaries limit severely the use of polycrystalline eucryptite for widespread ionic conduction applications. From a fundamental point of view, the use of one average conductivity for all slow pathways enabled us to determine three thermal regimes for the 1-dimensional ionic conduction (in bulk, parallel to the $c$ axis): the low temperature regime (below 440 $^o$C), order-disorder transformation regime (440-500 $^o$C), and the high temperature regime (above 500 $^o$C). Each of these regimes are well described by an Arhenius equation, and the calculations of activation energies coupled with molecular statics calculations of the potential energy surface shows that before and after the order-disorder transition the mechanism of Li conduction is correlated hopping. Within the narrow temperature range of the phase transformation (440-500 $^o$C), the calculated barriers are consistent with uncorrelated, individual hopping of Li atoms which occurs likely because in this temperature range the Li atoms have not yet assumed (on average) equispaced positions, and the correlation between hops along the same channel is broken by closely spaced ions.

## Acknowledgments

We gratefully acknowledge the support of U.S. Department of Energy's Office of Basic Energy Sciences through Grant No. DE-FG02-07ER46397, as well the partial support from the National Science Foundation through Grant No. DMR-1534503. We thank Dr. Ryan O'Hayre for insightful discussions, and Mr. Chuancheng Duan for his help in designing some of the experiments.

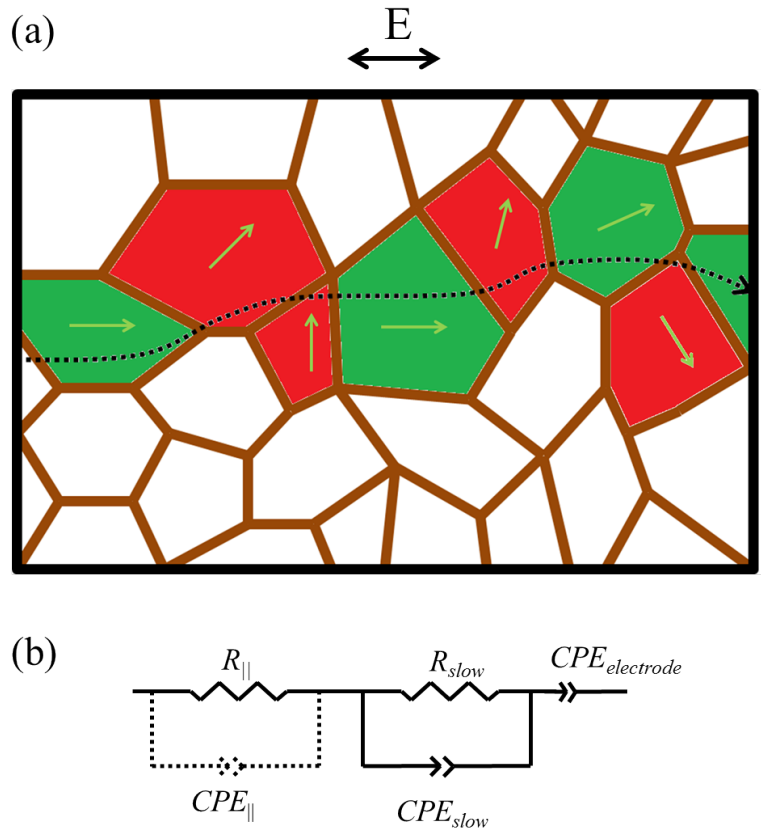

**Figure 1.** (a) Schematic drawing of polycrystalline β-eucryptite, with one current pathway drawn through the structure. The green arrows represent the direction of the *c*-axis, whose orientation with respect to the pathway renders some grains fast (green, pathway approximately along the *c* axis), and some grains slow (red, where the orientation is not sufficiently aligned with the pathway). (b) Equivalent circuit to interpret the complex impedance spectra, where $R_\parallel$ is the bulk resistance for conduction parallel to the *c* axis; $R_{slow}$ and $CPE_{slow}$ are the resistance and constant-phase element capacitance of the possible slow paths, including grain boundaries and unfavorably oriented grains; $CPE_{electrode}$, capacitance of the electrode. The constant-phase element $CPE_\parallel$ is represented as dotted lines since even when present, it cannot be detected from the experiments.

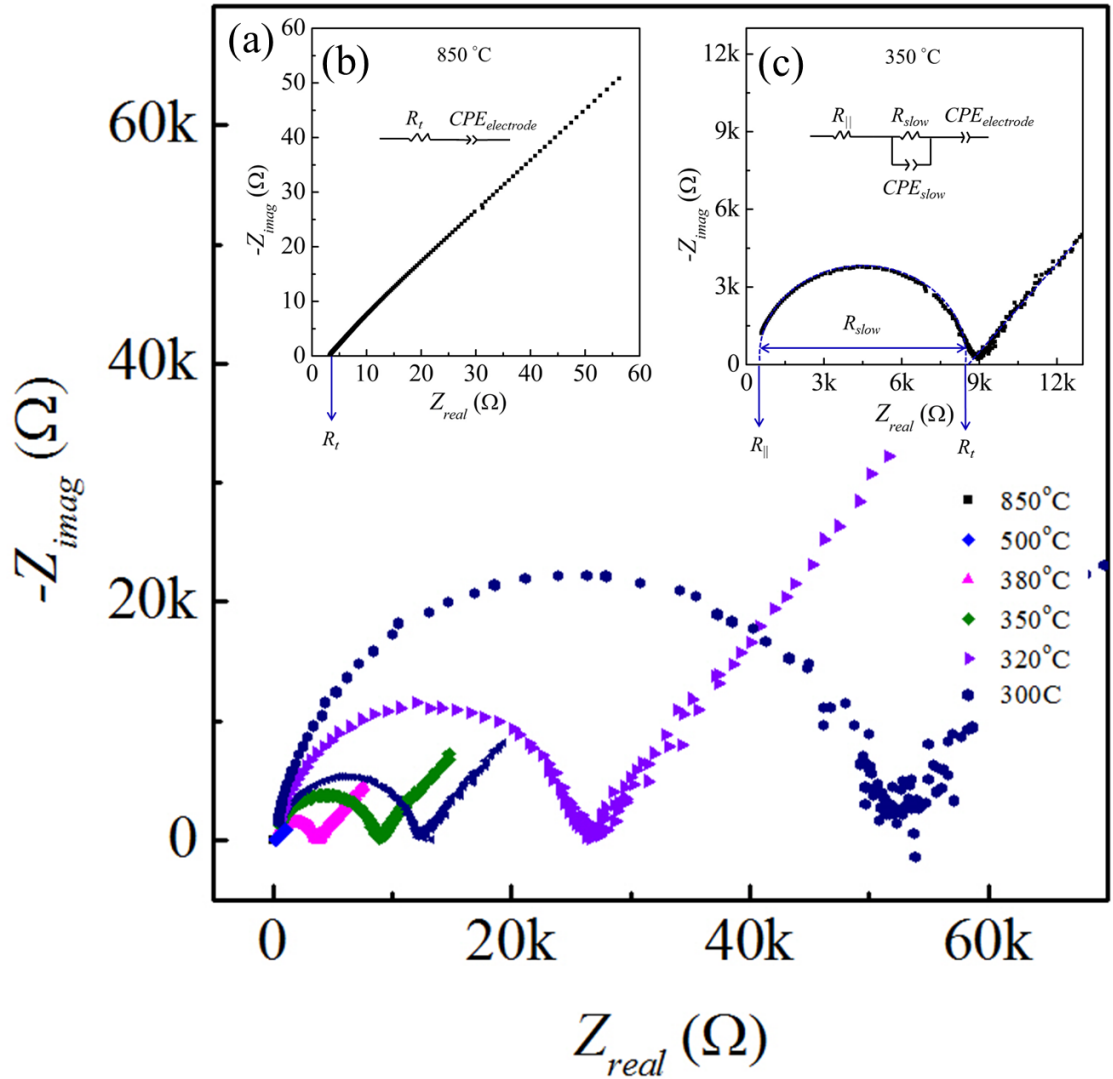

**Figure 2.** (a) Impedance spectra of pure β-eucryptite at selected temperatures from 300 to 850°C. (b, c) Impedance spectra and equivalent circuits at (b) 850 °C and (c) 350 °C.

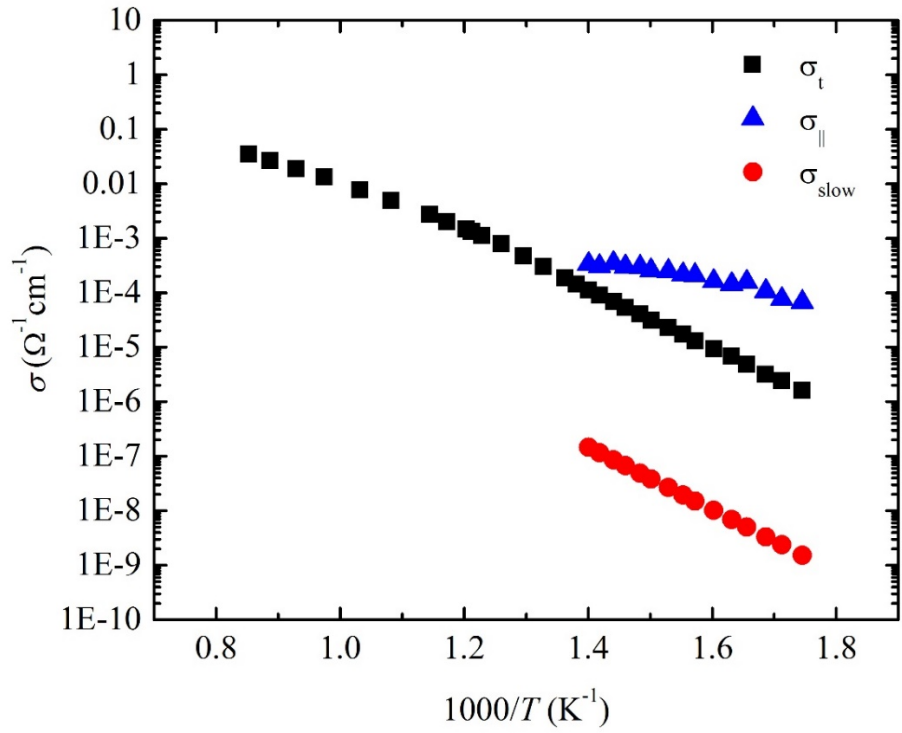

**Figure 3.** Total conductivity ($\sigma_t$), conductivity along the *c*-axis grains ($\sigma_\parallel$), and along slow paths ($\sigma_{slow}$) as functions of temperature.

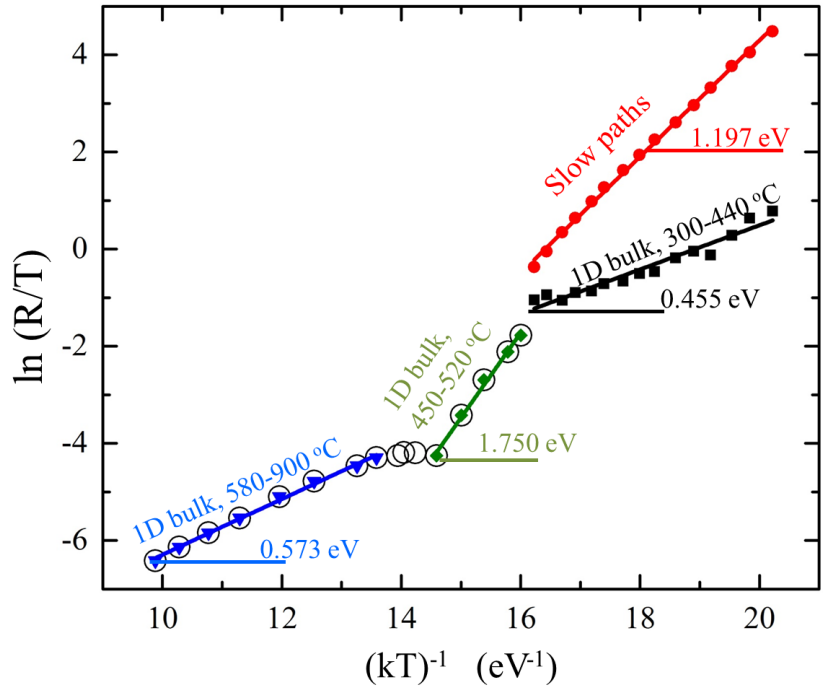

**Figure 4.** Experimentally measured ln(*R/T*) as a function of 1/*kT* for total, fast grains and slow paths at low temperatures (solid symbols) combined with calculated values at high temperatures (circled symbols). The activation energies are marked as the slopes of the linear portions.

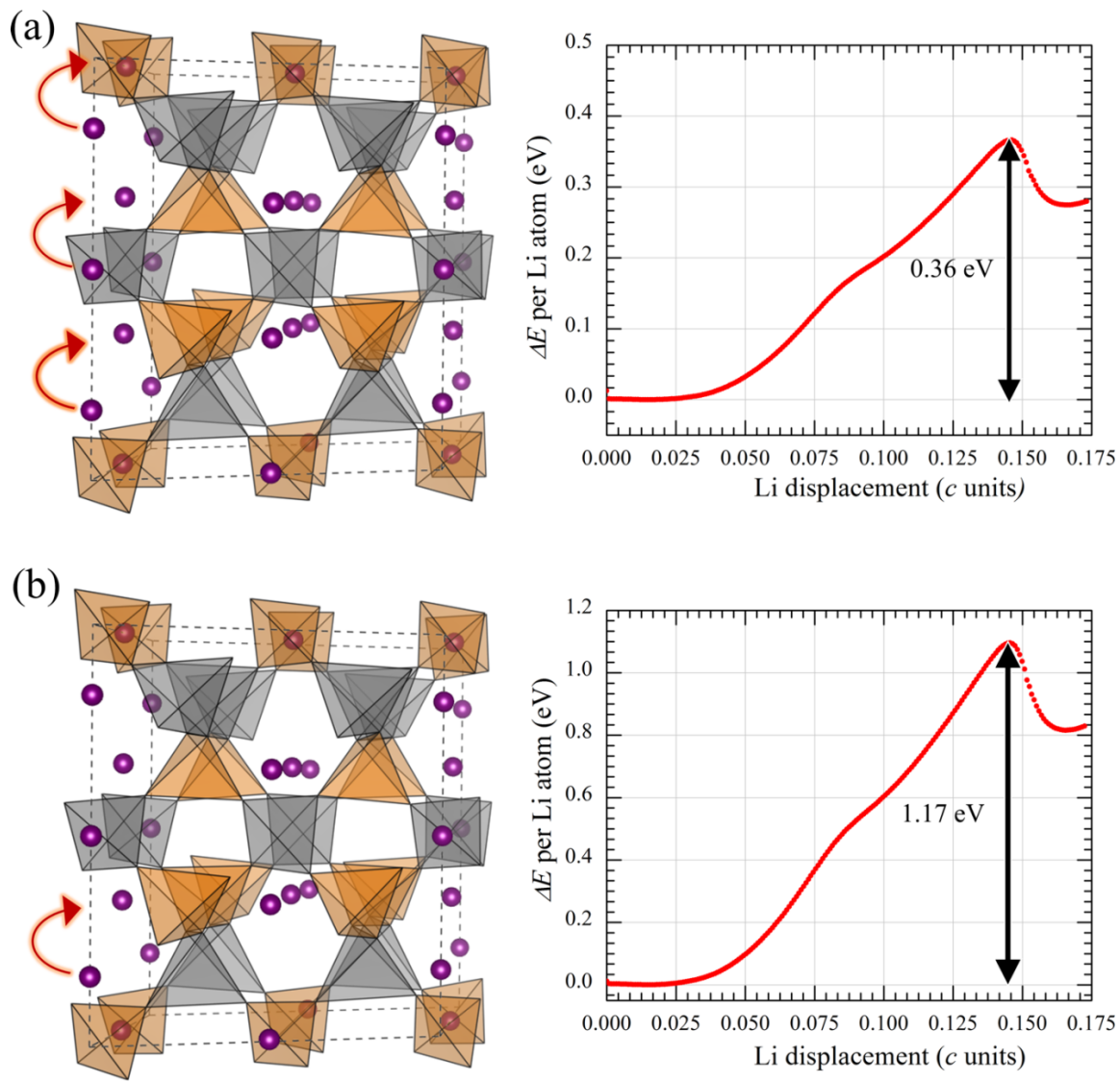

**Figure 5.** Side view of the β-eucryptite structures, with vertical channels in which Li motion can be correlated (a) or uncorrelated (b). Energy variation along the channels and the corresponding activation barrier for moving (a) three Li atoms at a time (correlated), and (b) one Li atom at a time (uncorrelated).